\documentclass{article}

\newcommand{\iu}{\texttt{j}}

\PassOptionsToPackage{numbers, compress}{natbib}
    \usepackage[preprint]{tackling_climate_workshop_style}

\usepackage{color}
\usepackage{xcolor}
\usepackage{amsmath}
\usepackage{xurl}
\usepackage{graphicx}
\usepackage{float}

\usepackage[utf8]{inputenc} % allow utf-8 input
\usepackage[T1]{fontenc}    % use 8-bit T1 fonts
\usepackage{hyperref}       % hyperlinks
\usepackage{url}            % simple URL typesetting
\usepackage{booktabs}       % professional-quality tables
\usepackage{amsfonts}       % blackboard math symbols
\usepackage{nicefrac}       % compact symbols for 1/2, etc.
\usepackage{microtype}      % microtypography

\definecolor{PeriwinkleGray}{rgb}{0.749,0.839,0.909}
\definecolor{PastelGreen}{rgb}{0.419,0.901,0.603}
\definecolor{PetiteOrchid}{rgb}{0.85,0.588,0.588}

\title{Optimal Power Grid Operations with Foundation Models}

\author{Alban Puech$^1$, Jonas Weiss$^1$, Thomas Brunschwiler$^1$, Hendrik F. Hamann$^2$ 
\\ $^1$IBM Research Europe, Switzerland
\\ $^2$IBM Research, T.J. Watson Research Center
} % TBD

\begin{document}

\maketitle

\begin{abstract} The energy transition, crucial for tackling the climate crisis, demands integrating numerous distributed, renewable energy sources into existing grids. Along with climate change and consumer behavioral changes, this leads to changes and variability in generation and load patterns, introducing significant complexity and uncertainty into grid planning and operations. While the industry has already started to exploit AI to overcome computational challenges of established grid simulation tools, we propose the use of AI Foundation Models (FMs) and advances in Graph Neural Networks to efficiently exploit poorly available grid data for different downstream tasks, enhancing grid operations. 
For capturing the grid's underlying physics, we believe that building a self-supervised model learning the power flow dynamics is a critical first step towards developing an FM for the power grid. We show how this approach may close the gap between the industry needs and current grid analysis capabilities, to bring the industry closer to optimal grid operation and planning.%, with associated gains in efficiency.
\end{abstract}

\vspace{2cm}
\section{Introduction and Problem Definition} % Why problem is important to climate change)
\textit{What challenges is the grid facing and why is solving them essential to tackling climate change?}

In 2018, global electric grid losses led to around 1 gigaton of CO$_2$ emissions, comparable to the annual emissions of half the U.S. households \cite{iea2020sustainable}. While optimizing grid operations is challenging due to its size and uncertainties, it is crucial to reduce these losses and fight climate change \cite{doe2022cleanelectricity}. 

The grid is divided into high-voltage transmission and lower-voltage distribution. Transmission operators manage four key tasks \cite{bose2021gridoperations}: \textit{grid operation} (ensuring operation within safe voltage and power limits), \textit{contingency analysis} (assessing resilience against outages), \textit{economic dispatch} (optimizing power generation costs), and \textit{capacity expansion} (increasing transmission capacity to meet future demand). 
A number of concurrent factors are adding uncertainty and complexity to grid operations \cite{hamann2024perspectivefoundationmodelselectric}: 
(1) The energy transition and the shift to weather-dependent renewable energy resources increase the variability in energy production patterns. 
(2) Changes in consumer behavior and accelerated electrification of the industry and mobility sectors affect demand profiles. Consequently, stochastic optimizations now need to run more scenarios, which is computationally more expensive. 
(3) The integration of distributed energy resources adds to the challenge, transforming the grid from a top-down system to a distributed network. This requires novel co-simulation frameworks for optimal power flow \cite{en14010012}, especially in distribution grids. 
(4) Extreme weather, aging infrastructure, and grid expansion further intensify the need for frequent and comprehensive contingency analyses \cite{BROCKWAY2020100256}.

Dealing with the added uncertainty and complexity requires larger and more frequent power flow analysis and optimization. However, current iterative power flow methods scale quadratically with grid nodes \cite{4596138} and can no longer cope with the problem sizes. The energy transition is thus dependent on the availability of faster, more efficient computational approaches \cite{doe2020advancedtransmission}.

\section{Proposed Solution}
\textit{How can AI bridge the gap between computational limits and the need for faster grid analysis and optimization?}

\textbf{Foundation Models (FMs) as scalable, operational grid models.} AI has already demonstrated great value for power grid operation and optimization\footnote{\scriptsize The US Department of Energy identifies ``improving power system optimization'' and ``simulating disruption/disaster scenarios'' as key opportunities for AI \cite{benes2024ai}. Improving grid operation and planning is one of the top five priorities in the U.S. Innovation to Meet 2050 Climate Goals \cite{whitehouse2022climategoals} and a key action in the Ten Actions Toward 100\% Clean Electricity \cite{denholm2022cleanelectricity,doe2022cleanelectricity}} \cite{daniel2024ai, smartcities4020029}.
With AI, instead of solving complex physics equations, we shift computational cost from inference to model training, speeding up run-time performance.
By exploiting AI FMs \cite{bommasani2022opportunitiesrisksfoundationmodels}, pre-trained on large datasets, and fine-tuned on smaller, labeled datasets, we overcome the typical AI challenges of limited application space and lack of topology transferability. The latter is addressed by exploiting known properties of Graph Neural Networks (GNNs) \cite{gnnreview}, facilitating efficient model adaptation to different grid topologies \cite{gnnreviewpowergrid}. 
By fine-tuning the model to multiple down-stream tasks or sub-grids, pre-training investments can be amortized. FMs have the required skill to approximate complex physical systems, as already demonstrated in weather forecasting \cite{FourCastNet}, with 4–5 orders of magnitude speedups \cite{bi2023accurate}.

\textbf{Solving power flow as a proxy for the underlying grid physics.} 
FMs are typically pre-trained on a suitable reconstruction task in a self-supervised manner \cite{bommasani2022opportunitiesrisksfoundationmodels}. The better pre-training learns structures and dynamics exploited in the downstream task, the better the final model will perform \cite{HAN2021225}. We thus suggest using power flow as a pre-training task.
It (1) describes fundamental power grid physics (Ohm’s and Kirchhoff’s laws \cite{powerflow, formulations}), and (2) is at the core of power grid operations with economic dispatch, contingency analysis, or capacity expansion. 
With increasing grid sizes and variable grid conditions, these tasks need to be executed increasingly frequently. 
They can thus greatly benefit from AI-accelerated power flow solutions \cite{piloto2024canosfastscalableneural, benes2024ai}. A simplified version of the power flow problem can be modeled as follows: 
% Transmission grids are represented as an undirected graph $\mathcal{G}=(\mathcal{N},\mathcal{E})$, where the nodes are the "buses" (load or generation points), and the edges are the "branches" (transmission lines between them). Let \( j \) be the unit imaginary number. We denote the net power injection at bus \( i \) by \( S_i = p_i + jq_i \), where \( p_i \) is the active power and \( q_i \) is the reactive power. Similarly, the power flow from bus \( i \) to bus \( j \) is denoted by \( S_{i,j} = p_{i,j} + jq_{i,j} \). The voltage at bus \( i \) is represented as \( V_i = v_i e^{j\delta_i} \), where \( v_i \) is the voltage magnitude and \( \delta_i \) is the voltage angle. The current between buses \( i \) and \( j \) is denoted by \( I_{i,j} \). The line impedance is represented by \( Z_{i,j} \in \mathbb{C} \) and is a fixed grid parameter. The state variables that describe the electricity flow within the grid include \( V_i \), \( S_i \), \( I_{i,j} \), and \( S_{i,j} \). We equip $\mathcal{G} $ with an arbitrary edge orientation $\mathcal{E}^D$ and denote by $I^*_{i,j}$ the complex conjugate of $I_{i,j}$. The power flow equations are as follows:
% \begin{align}
%     S_i &= \sum_{(i,j) \in \mathcal{E}^D } S_{i,j} - \sum_{(k,i) \in \mathcal{E}} (S_{k,i} - Z_{k,i} |I_{k,i}|^2), & \forall i \in \mathcal{N} \\
%      I_{i,j} &= (V_i - V_j) / Z_{i,j}, & \forall (i,j) \in \mathcal{E} \\
%     S_{i,j} &= V_i \cdot I_{i,j}^*, & \forall (i,j) \in \mathcal{E} 
% \end{align}
We represent a transmission grid as a directed graph \( \mathcal{G} = (\mathcal{N}, \mathcal{E}) \), where nodes \(\mathcal{N}\) are buses (e.g. loads, generators, or slack buses, used to balance power flow by absorbing generation-consumption differences) and edges \(\mathcal{E}\) are branches (e.g. transmission lines). The net power injection at bus \( i \) is \( S_i = p_i + \iu q_i \), with \( p_i \) the active, \( q_i \) the reactive power, and $\iu=\sqrt{-1}$. The power flow from bus \( i \) to bus \( j \) is \( S_{i,j} = p_{i,j} + \iu q_{i,j} \). The voltage at bus \( i \) is \( V_i = v_i e^{\iu \delta_i} \), and the current between buses \( i \) and \( j \) is \( I_{i,j} \in \mathbb{C} \). The complex impedance \( Z_{i,j} \in \mathbb{C} \) is a fixed grid parameter. We denote by $I^*_{i,j}$ the conjugate of $I_{i,j}$. The complex-form power flow equations are:

\begin{equation}
\begin{aligned}
    S_i &= \sum_{(i,j) \in \mathcal{E}^D} S_{i,j} - \sum_{(k,i) \in \mathcal{E}} (S_{k,i} - Z_{k,i} |I_{k,i}|^2), & \forall i \in \mathcal{N}  \\
    I_{i,j} &= (V_i - V_j)/(Z_{i,j}) & \forall (i,j) \in \mathcal{E} \\
    S_{i,j} &= V_i \cdot I_{i,j}^* & \forall (i,j) \in \mathcal{E} \\
\end{aligned}
\tag{A}
\label{eq:power_flow}
\end{equation}
The real valued equivalent of (A) has \( 2|\mathcal{N}| + 4|\mathcal{E}| \) equations with \( 4|\mathcal{N}| + 4|\mathcal{E}| \) variables. Given the value of at least \( 2|\mathcal{N}| \) real variables, (\ref{eq:power_flow}) can be solved for the remaining variables. In the ``standard'' \textit{power flow problem} \cite{powerflow}, the known variables are: the active (generated) power \( p_i \)  and voltage magnitude \( v_i \) at the generator buses, the active and reactive (load) power \( p_i \) and \( q_i \) at the load buses and the voltage magnitude \( v_i \) and angle \( \delta_i \) at the slack bus.

\textbf{Building on recent data and modeling advances of the power flow community.} 
In recent years, large power-grid datasets have been made available, created by 
(1) collecting grid topology and measurement data, 
(2) generating varying synthetic load conditions for one grid model, and 
(3) using power flow solvers to produce a wide range of power flow solutions under different conditions. 
For example, the one from \cite{lovett2024opfdatalargescaledatasetsac} contains 300,000 solved optimal power flow problems for each of the 10 different grid topologies it uses, including solutions for N-1 contingency analysis\footnote{\scriptsize N-1 contingency analysis checks if a grid can handle the failure of any single component without major disruptions.}, with grid-sizes ranging from 14 to 14,000 buses. 
As it only contains optimal power flow data, it is unclear whether models trained on it will perform well under non-optimal power flow conditions. 
\cite{varbella2024powergraphpowergridbenchmark} provides a dataset for power flow under real load conditions. It is limited to smaller grid topologies (up to 118 buses) and is not suitable for contingency analysis. GNNs based models have already been used to learn power flow, e.g. in \cite{varbella2024powergraphpowergridbenchmark}. In the latter, they were trained in a supervised, rather than unsupervised scheme as would be expected for FMs. % "expected" is not very accurate here...I know
They also didn't include power flow equations in the loss function, limiting its use to applications with relaxed accuracy requirements \cite{Radziukynas2009}.

\section{Methodology and Implementation}
\textit{ How do we plan on building a self-supervised model that learns the power flow dynamics?} 

\begin{figure}[H]
    \centering
\includegraphics[width=1\linewidth]{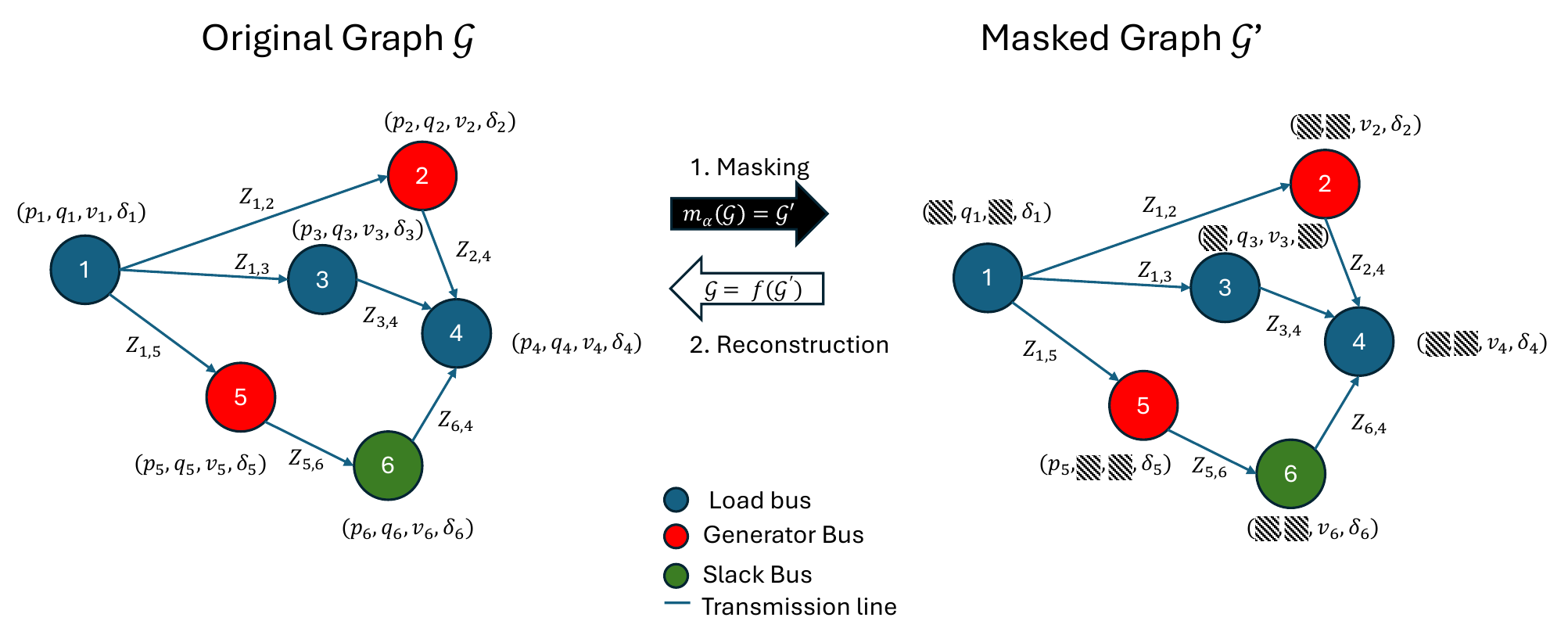}
    \caption{Masking and reconstruction steps for grid FM training. \textbf{1. Masking:} Given a graph representation of the transmission grid $\mathcal{G}$, the function $m_\alpha$ randomly masks node variables (independently of the bus type) with masking probability $\alpha$. The resulting masked graph is $\mathcal{G}'$. \textbf{2.  Reconstruction:} We assume the existence of a function $f$ that, given a masked graph $\mathcal{G}'$, returns the original graph $\mathcal{G}$. Our model is trained to approximate $f$.
    \label{fig:mae}}
\end{figure}

\textbf{Task and Architecture:} Let \( \mathcal{G} = (\mathcal{N}, \mathcal{E}) \) be a graph representation of a transmission grid, with node variables \( x_i = (p_i, q_i, v_i, \delta_i) \), for all \( i \in \mathcal{N} \) and edge parameters \( e_{i,j} = (Z_{i,j}) \) for all \( (i, j) \in \mathcal{E} \). \( S_{i,j} \) and \( I_{i,j} \) are not included in the graph, as they can be derived from node variables and edge parameters.
%\footnote{\scriptsize inferring node variables from known edge variable values is of limited interest for the downstream tasks we are currently interested in.} % I don't think we need this footnote ... 
To infer missing node variables from the known ones, we build a generative neural power flow equations solver. We train it as an autoencoder using self-supervised learning with node-feature masking strategies, such as GraphMAE \cite{graphmae, graphmae2}, on a reconstruction task, as shown in Figure \ref{fig:mae} in \ref{sub:mask}. Reconstructing a grid graph with all but \( 2|\mathcal{N}| \) variables masked effectively constitutes solving the power flow equations.

% We should better explain what components of the overall loss functions do and why they are there...is what I write correct?
\textbf{Loss:} The loss function is key to guide and constrain the model during reconstruction training towards meaningful node variable values and within physical boundaries. Within the loss \( \mathcal{L} = \mathcal{L}_{SCE} + \mathcal{L}_{power flow} \), the scaled cosine error \( \mathcal{L}_{SCE} \) minimizes the distance between the reconstructed and the original graphs \cite{graphmae}, while \( \mathcal{L}_{power flow} \) encourages satisfaction of the power flow equations by penalizing deviations between the left and right sides of the equations in (\ref{eq:power_flow}), similar to \cite{piloto2024canosfastscalableneural}.

\textbf{Data:} Our focus is reconstructing grid-graphs node variables. 
Initially, we will use the synthetic power flow dataset from \cite{varbella2024powergraphpowergridbenchmark}. 
In the following, to generate power flow problems with real-world profiles as in \cite{varbella2024powergraphpowergridbenchmark}, we will leverage graph topologies from PGLIB-OPF \cite{pglib}, and apply topology and load perturbations. 
These problems will be solved using Pandapower \cite{pandapower} and PowerModels.jl \cite{powermodels}. % Why those?
This will provide a collection of power flow solutions for various topologies, suitable for training the model on graphs with perturbed topologies (for contingency analysis) and real-world load profiles.

\textbf{Model Evaluation and Exploration:} 
In \cite{piloto2024canosfastscalableneural} and \cite{varbella2024powergraphpowergridbenchmark}, neural power flow solvers are topology-specific. We believe that a grid FM should also generalize well on diverse grid topologies, whether radial (e.g., Quebec Interconnection) or meshed (as often seen e.g. in the U.S.). 
We will assess and refine the model architecture and training data towards this goal. 
The key questions we will answer include: 
\textit{How well can a single model generalize to unseen topologies? % reformulated to "open question"
How many topology and parameter combinations are needed for training?
Should models be tailored to specific ranges of graph sizes? How much fine-tuning is required for optimal performance on a given topology?}

\section{Outlook and Expected Impact}
\textit{What immediate impacts can be unlocked by an FM, and what are the long-term prospects?}

% Outlook
\textbf{An FM model trained on power flow dynamics can be immediately applied to high-value tasks.} 
We expect minimal fine-tuning for downstream tasks that need power flow to be solved \textit{significantly faster}. With demonstrated speedups ranging from 2 (for a 500-bus grid) to 4 orders of magnitude (for a 10,000-bus grid) over traditional power flow solvers \cite{piloto2024canosfastscalableneural}, stochastic simulations of \textit{much larger grids} become practical. This will benefit applications like 
(1) replacing power flow solvers for tasks that don't require exact solutions e.g. instead of DC approximations \cite{wood1996power}; 
(2) functional/operational-aware power flow, if fine-tuned
with measured power flow data that contain functional and economic constraint.
(3) N-1 and higher order contingency analysis; or (4) operationally constrained optimal economic dispatch \cite{hamann2024perspectivefoundationmodelselectric}. The motivation and the implementation specifics are discussed in \ref{sub:down}. 

% Unclear what you mean with this!!:, also accounting for constraints not modeled by physics (e.g., losses).

% Expected Impact
\textbf{A grid FM has the prospect of becoming an early power grid digital twin.} 
Learning grid dynamics from real-time, operational grid measurements and adding operational constraints may help overcome traditional modeling uncertainties into a data-driven, more accurate representation of even very large grids. With the FM speedup advantage, more optimal grid scenarios can be found, leading to economic gains and higher efficiency, thus lower emission footprints. In Lithuania, e.g., AI models are monitoring and predicting line throughput, resulting in an increased usable transmission capacity of 52\% \cite{litgrid2024testing}. Large cost savings are also possible by mitigating power outages through optimal grid management. They are estimated to cost American businesses \$150 billion every year \cite{litos_smartgrid}. Exploiting the very large scenario handling and optimization capability of FMs has further the prospect of significantly impacting long-term grid expansion planning \cite{benes2024ai}.

% \textbf{Perspectives:}
% \begin{itemize}
% \item \textbf{Learning Grid Dynamics:} When trained on real data, the model is expected to learn grid dynamics more accurately than physical models, providing solutions closer to the true behavior of the grid.
% \item \textbf{Improved Optimization:} By better capturing grid dynamics, the model could yield solutions for optimization tasks like economic dispatch that are closer to the real optimum, as it accounts for losses and grid specificities that physical models do not.
% \item \textbf{Enhanced Scalability:} The model's speed and scalability could make tasks like power flow at the distribution level, bidirectional power flow, and transmission-distribution co-simulation more attainable.
% \item \textbf{Stochastic Optimization:} The model's efficiency enables more robust grid operation and planning by allowing the inclusion of more stochastic scenarios in optimization problems.
% \end{itemize}

\section*{Author Contributions}
Conceptualization, A.P. and J.W.; Investigation, A.P.; Writing –
Original Draft, A.P.; Writing – Editing, A.P. and J.W.; Writing – Review, T.B. and H.F.H.;

\section*{Acknowledgements}The authors express their sincere gratitude to Johannes Jakubik for his insightful comments on the paper.

\bibliographystyle{ieeetr}

\bibliography{bibliography}

\begin{thebibliography}{10}

\bibitem{iea2020sustainable}
{International Energy Agency}, ``Sustainable recovery,'' 2020.

\bibitem{doe2022cleanelectricity}
P.~Donohoo-Vallett, N.~Ryan, and R.~Wiser, ``On the path to 100\% clean electricity.'' U.S. Department of Energy, 2022.

\bibitem{bose2021gridoperations}
A.~Bose and T.~J. Overbye, ``Electricity transmission system research and development: Grid operations,'' 2021.
\newblock In Transmission Innovation Symposium: Modernizing the U.S. Electrical Grid, U.S. Department of Energy.

\bibitem{hamann2024perspectivefoundationmodelselectric}
H.~F. Hamann, T.~Brunschwiler, B.~Gjorgiev, L.~S.~A. Martins, A.~Puech, A.~Varbella, J.~Weiss, J.~Bernabe-Moreno, A.~B. Massé, S.~Choi, I.~Foster, B.-M. Hodge, R.~Jain, K.~Kim, V.~Mai, F.~Mirallès, M.~D. Montigny, O.~Ramos-Leaños, H.~Suprême, L.~Xie, E.-N.~S. Youssef, A.~Zinflou, A.~J. Belvi, R.~J. Bessa, B.~P. Bhattari, J.~Schmude, and S.~Sobolevsky, ``A perspective on foundation models for the electric power grid,'' 2024.

\bibitem{en14010012}
H.~Jain, B.~A. Bhatti, T.~Wu, B.~Mather, and R.~Broadwater, ``Integrated transmission-and-distribution system modeling of power systems: State-of-the-art and future research directions,'' {\em Energies}, vol.~14, no.~1, 2021.

\bibitem{BROCKWAY2020100256}
A.~M. Brockway and L.~N. Dunn, ``Weathering adaptation: Grid infrastructure planning in a changing climate,'' {\em Climate Risk Management}, vol.~30, p.~100256, 2020.

\bibitem{4596138}
P.~J. Lagace, M.~H. Vuong, and I.~Kamwa, ``Improving power flow convergence by newton raphson with a levenberg-marquardt method,'' in {\em 2008 IEEE Power and Energy Society General Meeting - Conversion and Delivery of Electrical Energy in the 21st Century}, pp.~1--6, 2008.

\bibitem{doe2020advancedtransmission}
{United States Department of Energy}, ``Advanced transmission technologies.'' \url{https://www.energy.gov/sites/prod/files/2021/02/f82/Advanced%20Transmission%20Technologies%20Report%20-%20final%20as%20of%2012.3%20-%20FOR%20PUBLIC.pdf}, December 2020.
\newblock Washington, DC 20585.

\bibitem{benes2024ai}
K.~J. Benes, J.~E. Porterfield, and C.~Yang, ``{AI for Energy: Opportunities for a Modern Grid and Clean Energy Economy},'' 2024.

\bibitem{whitehouse2022climategoals}
T.~W. House, ``U.s. innovation to meet 2050 climate goals: Assessing initial r\&d opportunities.'' \url{https://energy.mit.edu/wp-content/uploads/2023/02/US-Innovation-to-Meet-2050-Climate-Goals.pdf}, November 2022.

\bibitem{denholm2022cleanelectricity}
P.~Denholm, P.~Brown, W.~Cole, {\em et~al.}, ``Examining supply-side options to achieve 100\% clean electricity by 2035.'' \url{https://www.nrel.gov/docs/fy22osti/81644.pdf}, 2022.
\newblock Golden, CO: National Renewable Energy Laboratory. NREL/TP-6A40-81644.

\bibitem{daniel2024ai}
C.~Daniel, J.~C. Gehin, K.~Laurin-Kovitz, B.~Morreale, R.~Stevens, and W.~Tumas, ``Advanced research directions on ai for energy: Report on winter 2023 workshops.'' \url{https://www.anl.gov/ai/reference/ai-for-energy-report-2024}, 2024.

\bibitem{smartcities4020029}
O.~A. Omitaomu and H.~Niu, ``Artificial intelligence techniques in smart grid: A survey,'' {\em Smart Cities}, vol.~4, no.~2, pp.~548--568, 2021.

\bibitem{bommasani2022opportunitiesrisksfoundationmodels}
R.~Bommasani, D.~A. Hudson, E.~Adeli, R.~Altman, S.~Arora, S.~von Arx, M.~S. Bernstein, J.~Bohg, A.~Bosselut, E.~Brunskill, E.~Brynjolfsson, S.~Buch, D.~Card, R.~Castellon, N.~Chatterji, A.~Chen, K.~Creel, J.~Q. Davis, D.~Demszky, C.~Donahue, M.~Doumbouya, E.~Durmus, S.~Ermon, J.~Etchemendy, K.~Ethayarajh, L.~Fei-Fei, C.~Finn, T.~Gale, L.~Gillespie, K.~Goel, N.~Goodman, S.~Grossman, N.~Guha, T.~Hashimoto, P.~Henderson, J.~Hewitt, D.~E. Ho, J.~Hong, K.~Hsu, J.~Huang, T.~Icard, S.~Jain, D.~Jurafsky, P.~Kalluri, S.~Karamcheti, G.~Keeling, F.~Khani, O.~Khattab, P.~W. Koh, M.~Krass, R.~Krishna, R.~Kuditipudi, A.~Kumar, F.~Ladhak, M.~Lee, T.~Lee, J.~Leskovec, I.~Levent, X.~L. Li, X.~Li, T.~Ma, A.~Malik, C.~D. Manning, S.~Mirchandani, E.~Mitchell, Z.~Munyikwa, S.~Nair, A.~Narayan, D.~Narayanan, B.~Newman, A.~Nie, J.~C. Niebles, H.~Nilforoshan, J.~Nyarko, G.~Ogut, L.~Orr, I.~Papadimitriou, J.~S. Park, C.~Piech, E.~Portelance, C.~Potts, A.~Raghunathan, R.~Reich, H.~Ren, F.~Rong, Y.~Roohani, C.~Ruiz, J.~Ryan, C.~Ré,
  D.~Sadigh, S.~Sagawa, K.~Santhanam, A.~Shih, K.~Srinivasan, A.~Tamkin, R.~Taori, A.~W. Thomas, F.~Tramèr, R.~E. Wang, W.~Wang, B.~Wu, J.~Wu, Y.~Wu, S.~M. Xie, M.~Yasunaga, J.~You, M.~Zaharia, M.~Zhang, T.~Zhang, X.~Zhang, Y.~Zhang, L.~Zheng, K.~Zhou, and P.~Liang, ``On the opportunities and risks of foundation models.'' \url{https://arxiv.org/abs/2108.07258}, 2022.

\bibitem{gnnreview}
J.~Zhou, G.~Cui, S.~Hu, Z.~Zhang, C.~Yang, Z.~Liu, L.~Wang, C.~Li, and M.~Sun, ``Graph neural networks: A review of methods and applications,'' {\em AI Open}, vol.~1, pp.~57--81, 2020.

\bibitem{gnnreviewpowergrid}
W.~Liao, B.~Bak-Jensen, J.~Pillai, Y.~Wang, and Y.~Wang, ``A review of graph neural networks and their applications in power systems,'' {\em Journal of Modern Power Systems and Clean Energy}, vol.~10, pp.~345--360, Mar. 2022.

\bibitem{FourCastNet}
J.~Pathak, S.~Subramanian, P.~Harrington, S.~Raja, A.~Chattopadhyay, M.~Mardani, T.~Kurth, D.~Hall, Z.~Li, K.~Azizzadenesheli, P.~Hassanzadeh, K.~Kashinath, and A.~Anandkumar, ``Fourcastnet: A global data-driven high-resolution weather model using adaptive fourier neural operators.'' \url{https://arxiv.org/abs/2202.11214}, 2022.

\bibitem{bi2023accurate}
K.~Bi, L.~Xie, H.~Zhang, X.~Chen, X.~Gu, and Q.~Tian, ``Accurate medium-range global weather forecasting with 3d neural networks,'' {\em Nature}, vol.~619, pp.~1--6, 07 2023.

\bibitem{HAN2021225}
X.~Han, Z.~Zhang, N.~Ding, Y.~Gu, X.~Liu, Y.~Huo, J.~Qiu, Y.~Yao, A.~Zhang, L.~Zhang, W.~Han, M.~Huang, Q.~Jin, Y.~Lan, Y.~Liu, Z.~Liu, Z.~Lu, X.~Qiu, R.~Song, J.~Tang, J.-R. Wen, J.~Yuan, W.~X. Zhao, and J.~Zhu, ``Pre-trained models: Past, present and future,'' {\em AI Open}, vol.~2, pp.~225--250, 2021.

\bibitem{powerflow}
M.~Farivar and S.~H. Low, ``Branch flow model: Relaxations and convexification—part i,'' {\em IEEE Transactions on Power Systems}, vol.~28, no.~3, pp.~2554--2564, 2013.

\bibitem{formulations}
A.~Mary, B.~Cain, and R.~O’Neill, ``History of optimal power flow and formulations,'' {\em Fed. Energy Regul. Comm.}, vol.~1, pp.~1--36, 01 2012.

\bibitem{piloto2024canosfastscalableneural}
L.~Piloto, S.~Liguori, S.~Madjiheurem, M.~Zgubic, S.~Lovett, H.~Tomlinson, S.~Elster, C.~Apps, and S.~Witherspoon, ``Canos: A fast and scalable neural ac-opf solver robust to n-1 perturbations.'' \url{https://arxiv.org/abs/2403.17660}, 2024.

\bibitem{lovett2024opfdatalargescaledatasetsac}
S.~Lovett, M.~Zgubic, S.~Liguori, S.~Madjiheurem, H.~Tomlinson, S.~Elster, C.~Apps, S.~Witherspoon, and L.~Piloto, ``Opfdata: Large-scale datasets for ac optimal power flow with topological perturbations.'' \url{https://arxiv.org/abs/2406.07234}, 2024.

\bibitem{varbella2024powergraphpowergridbenchmark}
A.~Varbella, K.~Amara, B.~Gjorgiev, M.~El-Assady, and G.~Sansavini, ``Powergraph: A power grid benchmark dataset for graph neural networks.'' \url{https://arxiv.org/abs/2402.02827}, 2024.

\bibitem{Radziukynas2009}
V.~Radziukynas and I.~Radziukyniene, {\em Optimization Methods Application to Optimal Power Flow in Electric Power Systems}, pp.~409--436.
\newblock Berlin, Heidelberg: Springer Berlin Heidelberg, 2009.

\bibitem{graphmae}
Z.~Hou, X.~Liu, Y.~Cen, Y.~Dong, H.~Yang, C.~Wang, and J.~Tang, ``Graphmae: Self-supervised masked graph autoencoders.'' \url{https://arxiv.org/abs/2205.10803}, 2022.

\bibitem{graphmae2}
Z.~Hou, Y.~He, Y.~Cen, X.~Liu, Y.~Dong, E.~Kharlamov, and J.~Tang, ``Graphmae2: A decoding-enhanced masked self-supervised graph learner,'' in {\em Proceedings of the ACM Web Conference 2023}, WWW '23, (New York, NY, USA), p.~737–746, Association for Computing Machinery, 2023.

\bibitem{pglib}
S.~Babaeinejadsarookolaee, A.~Birchfield, R.~D. Christie, C.~Coffrin, C.~DeMarco, R.~Diao, M.~Ferris, S.~Fliscounakis, S.~Greene, R.~Huang, C.~Josz, R.~Korab, B.~Lesieutre, J.~Maeght, T.~W.~K. Mak, D.~K. Molzahn, T.~J. Overbye, P.~Panciatici, B.~Park, J.~Snodgrass, A.~Tbaileh, P.~V. Hentenryck, and R.~Zimmerman, ``The power grid library for benchmarking ac optimal power flow algorithms,'' 2021.

\bibitem{pandapower}
L.~Thurner, A.~Scheidler, F.~Sch{\"a}fer, J.~Menke, J.~Dollichon, F.~Meier, S.~Meinecke, and M.~Braun, ``pandapower — an open-source python tool for convenient modeling, analysis, and optimization of electric power systems,'' {\em IEEE Transactions on Power Systems}, vol.~33, pp.~6510--6521, Nov 2018.

\bibitem{powermodels}
C.~Coffrin, R.~Bent, K.~Sundar, Y.~Ng, and M.~Lubin, ``Powermodels.jl: An open-source framework for exploring power flow formulations,'' in {\em 2018 Power Systems Computation Conference (PSCC)}, pp.~1--8, June 2018.

\bibitem{wood1996power}
A.~J. Wood and B.~F. Wollenberg, {\em Power Generation, Operation, and Control}.
\newblock John Wiley \& Sons, 2nd~ed., 1996.

\bibitem{litgrid2024testing}
{Litgrid}, ``{Litgrid} has completed a year of testing with an average 52 percent increase in transmission line capacity.'' \url{https://www.litgrid.eu/index.php/naujienos/naujienos/-litgrid-baige-metus-trukusius-bandymus-elektros-perdavimo-linijos-pralaidumas-vidutiniskai-padidejo-52-procentais/32687}, January 2024.
\newblock Original title in Lithuanian: „Litgrid“ baigė metus trukusius bandymus: elektros perdavimo linijos pralaidumas vidutiniškai padidėjo 52 procentais.

\bibitem{litos_smartgrid}
{Litos Strategic Communication}, ``The smart grid: An introduction.'' \url{http://energy.gov/sites/prod/files/oeprod/DocumentsandMedia/DOE_SG_Book_Single_Pages(1).pdf}, 2008.
\newblock U.S. Department of Energy.

\bibitem{4073219}
W.~F. Tinney and C.~E. Hart, ``Power flow solution by newton's method,'' {\em IEEE Transactions on Power Apparatus and Systems}, vol.~PAS-86, no.~11, pp.~1449--1460, 1967.

\bibitem{7798623}
K.~Dvijotham and D.~K. Molzahn, ``Error bounds on the dc power flow approximation: A convex relaxation approach,'' in {\em 2016 IEEE 55th Conference on Decision and Control (CDC)}, pp.~2411--2418, 2016.

\bibitem{infeasible}
K.~Baker, ``Solutions of dc opf are never ac feasible,'' in {\em Proceedings of the Twelfth ACM International Conference on Future Energy Systems}, e-Energy '21, (New York, NY, USA), p.~264–268, Association for Computing Machinery, 2021.

\end{thebibliography}

\clearpage

\section{Appendix}

\subsection{Masking and reconstruction steps}
\label{sub:mask}

\subsection{Downstream tasks}
\label{sub:down}

We provide an overview of the motivation and implementation directions for a selection of high-value downstream tasks that directly leverage our proposed model:

(1) \textbf{Replacing Numerical Power Flow Solvers.} \textit{Motivation:} Solving power flow problems is computationally expensive, with the most accurate solvers relying on Newton's method to iteratively solve the AC power flow equations \cite{4073219} (refer to Equations (\ref{eq:power_flow})). \textit{Implementation:} 
For applications that require solving power flow many times, where speed is valued over accuracy, linearized approximations (such as DC power flow \cite{7798623}) are typically used. 
These applications can directly benefit from faster AI-based power flow solvers, used in place of DC power flow solvers. 
We expect minimal fine-tuning of our model for power flow solving, as the task is equivalent to the suggested pre-training reconstructing task when masking all node variables except for: active (generated) power \( p_i \) and voltage magnitude \( v_i \) at the generator buses, active and reactive (load) power \( p_i \) and \( q_i \) at the load buses, and the voltage magnitude \( v_i \) and angle \( \delta_i \) at the slack bus. Depending on the masking strategy during training, if sufficient number of graphs are masked according to the power flow problem, no fine-tuning may be necessary at all.

(2) \textbf{Functional/Operational-Aware Power Flow.} \textit{Motivation:} While power flow analysis is essential for grid modeling and analysis, it does not account for important functional constraints or economic factors inherent in grid operation. 
This is problematic for applications like operational planning, capacity expansion (where various grid configurations are simulated for future load and generation conditions), or contingency analysis (where it is crucial to determine whether the grid can handle the loss of a generator or line). 
Accurate representation of grid operation for these scenarios is vital, but it is difficult to model using traditional mathematical approaches. 
However, these dynamics can be inferred and learned by data-driven AI models from observed and historic grid operations data, enabling more accurate grid simulations, improving grid planning, and potentially reducing over-provisioning and safety margins. \textit{Implementation:} After pre-training on a collection of power flow problems solved using physics-based methods, our model learns the power flow dynamics as described by the equations. 
Complementing this initial pre-training with a second, self-supervised reconstruction phase using real data collected under various conditions over an extended period of time should augment the model with the additional dynamics associated with grid operation. % This is quite a tricky statement ... as many operational data depend on too many variables, weather, social behaviour etc. such that I am a little sceptical here. 
Another possibility is to use the pre-trained model to generate grid embeddings, which could then be used in conjunction with another model that accounts for external information such as time of the day, date, and other factors typically influencing grid operation. This combined model could be fine-tuned on real-world data to capture operational nuances.

(3) \textbf{Contingency Analysis.} \textit{Motivation:} Contingency analysis is essential for assessing grid resilience. In an N-1 contingency analysis, the loss of one line, one transformer, or one generator is simulated, and grid stability is assessed for many possible scenarios using power flow analysis.
Currently, operators are constrained by the time required to perform a single power flow analysis, limiting the number of scenarios they can evaluate. 
For example, an N-1 contingency analysis on the Western Interconnection takes approximately 5 minutes \cite{benes2024ai}. On a grid with 1,000 lines, simulating all possible losses of 1 line requires solving 1,000 power flow problems. However, simulating scenarios involving the losses of any 2 components, requires 499,500 power flow analyses, as the problem scales with \(k\) (the number of lost elements) as \(O(N^k)\), where \(N\) is the number of candidate lost elements. A 100 times faster power flow solver would allow operators to run contingency analyses with many more scenarios (solving 499,500 scenarios with AI would only take as long as solving ~ 5,000 scenarios using traditional physics-based power flow solvers), thereby greatly enhancing grid resilience assessments.
\textit{Implementation:} For a power flow model that additionally accounts for changing grid topologies, experiments will be necessary to determine if the model can directly manage this extra complexity. 
Alternatively a fine-tuned model may be used, on datasets obtained by solving power flow problems on grids with perturbed topologies (e.g., with dropped generators, lines, transformers, etc.).

(4) \textbf{Operationally Constrained Optimal Dispatch.} \textit{Motivation:}  The goal of optimal power flow (OPF) is to determine the most efficient operating conditions for a power system. Economic dispatch is a variant of OPF that specifically aims to minimize generation costs. OPF problems are typically solved using interior point methods, which require the computation of the Hessian of a Lagrangian function at each step. Accelerating the solution of power flow problems while integrating both functional and economic operator constraints is of high value. This can replace DC-approximation-based solvers, that may give infeasible solutions \cite{infeasible}. For example, \cite{piloto2024canosfastscalableneural} demonstrated an OPF speedup of 2 to 4 orders of magnitude and accurate solutions, but the model was trained on synthetic data only. \textit{Implementation:} OPF is an optimization problem, distinct from the standard power flow problem. The generator outputs are unknown and must be determined to meet load demands, while respecting generator capacities, and line capacity constraints. Whether OPF can be treated as a reconstruction task by masking the generator power in addition to the variables usually masked in power flow problems is subject of further studies. 
The challenge is that the reconstructed generator power must also minimize an objective function (like cost in economic dispatch), which the model does not know. We anticipate good performance of models fine-tuned on reconstruction tasks with fixed generator cost, but they may not generalize well to other cost functions or generator capacity constraints. 
The challenge is exacerbated by reconstructed node variables that must adhere to bounds (e.g., generator capacity constraints) that the model does not know either, calling into question a feasible solution altogether.
However, we are confident that power flow solvers can be hierarchically integrated into optimization schemes where generator power is iteratively adjusted, solution feasibility is checked, and other variables are determined using the power flow solver. 
By combining existing optimization schemes to modify the generator variables, an optimal setting within the solution space can then be found. This is an approach we will explore in the future.

\end{document}